\documentstyle[12pt,epsfig]{article}
\begin{document}
\centerline{\Large\bf Decrumpling or TVSD model
explains why the universe}
\vspace*{0.050truein}
\centerline{\Large\bf is accelerating today}
\vspace*{0.050truein}
\centerline{Forough Nasseri\footnote{Email: nasseri@fastmail.fm}}
\centerline{\it Physics Department, Sabzevar University of Tarbiat
Moallem, P.O.Box 397, Sabzevar, Iran}
\begin{center}
(\today)
\end{center}
\begin{abstract}
Within the framework of a model universe with time variable space
dimension (TVSD), known as decrumpling or TVSD model, we show
the present value of the deceleration parameter is negative
implying that the universe is accelerating today.
Our study is based on a flat universe with the equation of state
parameter to be $\omega(z=0) \approx -1$ today. More clearly, decrumpling
model tells us the universe is accelerating today due to the cosmological
constant which is the simplest candidate for the dark energy.

\end{abstract}

In recent years the exploration of the universe at redshifts of order
unity has provided information about the time evolution of the expansion
rate of the universe. Observations indicate that the universe is
presently undergoing a phase of accelerated expansion \cite{1}.
The goal of this letter is to show a model universe with time
variable space dimension, known as decrumpling model, explains why
the universe is accelerating today.
Our approach to conclude this result is to compute
the deceleration parameter of the model and to show its present value  
is negative, implying the universe is accelerating today.
For more details about decrumpling or TVSD model see Refs. \cite{2}-\cite{9}.

We will use the natural units system that sets $k_B$, $c$, and $\hbar$
all equal to one, so that $\ell_P=M_P^{-1}=\sqrt{G}$.
To read easily this letter we also use the notation $D_t$ instead of
$D(t)$ that means the space dimension $D$ is as a function of time.

In Refs.\cite{2}-\cite{9}, the decrumpling or TVSD model has been
studied. Assume the universe consists of a fixed number $\cal{N}$ of
universal cells having a characteristic length $\delta$ in each of their
dimensions. The volume of the universe at the time $t$ depends
on the configuration of the cells. It is easily seen that 
\begin{equation}
\label{1}
{\rm vol}_{D_t}({\rm cell})={\rm vol}_{D_0}({\rm cell})\delta^{D_t-D_0},
\end{equation}
where the $t$ subscript in $D_t$ means that $D$ to be as a function
of time, i.e. $D(t)$. 

Interpreting the radius of the universe, $a$, as the radius of
gyration of a crumpled ``universal surface'',
the volume of space can be written 
\begin{eqnarray}
\label{2}
a^{D_t}&=&{\cal{N}}\,{\rm vol}_{D_t}({\rm cell})\nonumber\\
   &=&{\cal{N}}\,{\rm vol}_{D_0}({\rm cell}) \delta^{{D_t}-D_0}\nonumber\\
   &=&{a_0}^{D_0} \delta^{{D_t}-D_0}
\end{eqnarray}
or
\begin{equation}
\label{3}
\left( \frac{a}{\delta} \right)^{D_t}=
\left( \frac{a_0}{\delta} \right)^{D_0} = e^C,
\end{equation}
where $C$ is a universal positive constant. Its value has a strong
influence on the dynamics of spacetime, for example on the dimension
of space, say, at the Planck time. Hence, it has physical and cosmological
consequences and may be determined by observation. The zero subscript in
any quantity, e.g. in $a_0$ and $D_0$, denotes its present value.
We coin the above relation as a``dimensional constraint" which relates
the ``scale factor" of decrumpling model to the spatial dimension.
We consider the comoving length of the Hubble
radius at present time to be equal to one. So the interpretation of the
scale factor as a physical length is valid.
The dimensional constraint can be written in this form
\begin{equation}
\label{4}
\frac{1}{D_t}=\frac{1}{C}\ln \left( \frac{a}{a_0} \right) + \frac{1}{D_0}.
\end{equation}
It is seen that by the expansion of the universe, the space
dimension decreases.
Time derivative of (\ref{3}) or (\ref{4}) leads to
\begin{equation}
\label{5}
{\dot D}_t=-\frac{D_t^2 \dot{a}}{Ca}.
\end{equation}
It can be easily shown that the case of constant space dimension
corresponds to when $C$ tends to infinity. In other words,
$C$ depends on the number of fundamental cells. For $C \to +\infty$,
the number of cells tends to infinity and $\delta\to 0$.
In this limit, the dependence between the space dimensions and
the radius of the universe is removed, and consequently we
have a constant space dimension.

We define $D_{P}$ as the space dimension of the universe when
the scale factor is equal to the Planck length $\ell_{P}$.
Taking $D_0=3$ and the scale of the universe today to be the present
value of the Hubble radius $H_0^{-1}$ and the space dimension at the
Planck length to be $4, 10,$ or $25$, from Kaluza-Klein and superstring
theory, we can obtain from (\ref{3}) and (\ref{4}) the corresponding
value of $C$ and $\delta$
\begin{eqnarray}
\label{6}
\frac{1}{D_P}&=&\frac{1}{C} \ln \left( \frac{\ell_P}{a_0}
\right) + \frac{1}{D_0}
= \frac{1}{C} \ln \left( \frac{\ell_P}{{H_0}^{-1}} \right) +\frac{1}{3},\\
\label{7}
\delta &=& a_0 e^{-C/D_0}= H_0^{-1} e^{-C/3}.
\end{eqnarray}
In Table 1, values of $C$, $\delta$ and also ${\dot D}_t|_0$ for some
interesting values of $D_P$ are given.
These values are calculated by assuming $D_0=3$ and
$H_0^{-1}=3000 h_0^{-1} {\rm Mpc}=9.2503 \times 10^{27} h_0^{-1}{\rm cm}$,
where we take $h_0=1$.

\begin{table}
\caption{Values of $C$ and $\delta$ for some values of
$D_P$. Time variation of space dimension today
has been also calculated in terms of sec$^{-1}$ and
yr$^{-1}$.}
\begin{tabular}{ccccc} \\ \hline\hline 
$D_P$ & $C$   & $\delta$ (cm)   & $\dot D|_0$ (sec$^{-1}$)  & $\dot D|_0$ (yr$^{-1}$) \\ \hline\hline
$3$           & $ +\infty$         &  $0$           & $0$ & $0$ \\ \hline
$4$           & $1678.797$         &  $8.6158 \times 10^{-216}$  & $-1.7374 \times 10^{-20} h_0$ & $ -5.4827 \times 10^{-13} h_0$  \\ \hline
$10$          & $599.571$          &  $1.4771 \times 10^{-59}$  &  $-4.8648 \times 10^{-20} h_0$ & $ -1.5352 \times 10^{-12} h_0$  \\ \hline
$25$          & $476.931$          &  $8.3810 \times 10^{-42}$  & $-6.1158 \times 10^{-20} h_0$ & $-1.9299 \times 10^{-12} h_0$ \\ \hline
$+\infty$     & $419.699$          &  $\ell_P$  & $-6.9498 \times 10^{-20} h_0$ & $ -2.1931 \times 10^{-12}h_0$ \\ \hline\hline
\end{tabular}
\end{table}

Let us define the action of the model for the special
Friedmann-Robertson-Walker (FRW) metric in an arbitrary fixed space
dimension $D$, and then try to generalize it to variable dimension.
Now, take the metric in constant $D+1$ dimensions in the following
form
\begin{equation}
\label{8}
ds^2 = -N^2(t)dt^2+a^2(t)d\Sigma_k^2,
\end{equation}
where $N(t)$ denotes the lapse function and $d\Sigma_k^2$ is the line
element for a D-manifold of constant curvature $k = + 1, 0, - 1$. The
Ricci scalar is given by
\begin{equation}
\label{9}
R=\frac{D}{N^2}\left\{\frac{2\ddot a}{a}+(D-1)\left[\left(\frac{\dot a}{a}
\right)^2 + \frac{N^2k}{a^2}\right]-\frac{2\dot a\dot N}{aN}\right\}.
\end{equation}
Substituting from Eq.(\ref{9}) in the Einstein-Hilbert action for
pure gravity in constant $D+1$ dimensions
\begin{equation}
\label{10}
S_G = \frac{1}{2\kappa} \int d^{(1+D)} x \sqrt{-g}R,
\end{equation}
and using the Hawking-Ellis action of a perfect fluid
for the model, the following Lagrangian has been obtained \cite{9}
\begin{equation}
\label{11}
L_I := -\frac{V_{D_t}}{2 \kappa N} \left( \frac{a}{a_0} \right)^{D_t}
D_t(D_t-1)
\left[ \left( \frac{\dot a}{a} \right )^2 -\frac{N^2 k}{a^2} \right ]
- \rho N V_{D_t} \left( \frac{a}{a_0} \right )^{D_t},
\end{equation}
where $\kappa=8 \pi {M_P}^{-2}=8 \pi G$, $\rho$ the energy density,
and $V_{D_t}$ the volume of the space-like sections

\begin{eqnarray}
\label{12}
V_{D_t}&=&\frac{2 \pi^{(D_t+1)/2}}{\Gamma[(D_t+1)/2]},\;\;\mbox{closed universe, $k=+1$,}\\
\label{13}
V_{D_t}&=&\frac{\pi^{(D_t/2)}}{\Gamma(D_t/2+1)}{\chi_c}^{D_t},\;\;\mbox{flat universe, $k=0$,}\\
\label{14}
V_{D_t}&=&\frac{2\pi^{(D_t/2)}}{\Gamma(D_t/2)}f(\chi_c),\;\;\mbox{open universe, $k=-1$,}
\end{eqnarray}
where $\chi_C$ is a cut-off and $f(\chi_c)$ is a function thereof
(for more details see Ref. \cite{9}).
In the limit of constant space dimensions, or $D_t=D_0$,
$L_I$ approaches to the Einstein-Hilbert Lagrangian
which is
\begin{equation}
\label{15}
L_{I}^0 := - \frac{V_{D_0}}{2 \kappa_0 N}
\left( \frac{a}{a_0} \right)^{D_0} D_0(D_0-1)
\left[ \left( \frac{\dot{a}}{a} \right)^2 - \frac{N^2 k}{a^2} \right ]
- \rho N V_{D_0} \left( \frac{a}{a_0} \right )^{D_0},
\end{equation}
where $\kappa_0=8\pi G_0$ and the zero subscript in $G_0$ denotes its
present value. So, Lagrangian $L_I$ cannot abandon Einstein's gravity.
Varying the Lagrangian $L_I$ with respect to $N$ and $a$, we find the
following equations of motion in the gauge $N=1$, respectively
\begin{eqnarray}
\label{16}
&&\left( \frac{\dot a}{a} \right)^2 +\frac{k}{a^2} =
\frac{2 \kappa \rho}{D_t(D_t-1)},\\
\label{17}
&&(D_t-1) \bigg\{ \frac{\ddot{a}}{a} + \left[ \left( \frac{\dot a}{a}
\right)^2
+\frac {k}{a^2} \right]
\bigg(
-\frac{D_t^2}{2C} \frac{d \ln V_{D_t}}{d{D_t}}
-1-\frac{D_t(2D_t-1)}{2C(D_t-1)} 
+\frac{D_t^2}{2D_0} \bigg)
\bigg\} \nonumber\\
&&+ \kappa p \bigg( -\frac{d \ln V_{D_t}}{d{D_t}} \frac{D_t}{C} 
-\frac{D_t}{C} \ln \frac{a}{a_0} +1 \bigg) =0.
\end{eqnarray}
Using (\ref{16}) and (\ref{17}), the evolution equation of the space
dimension can be obtained by
\begin{equation}
\label{18}
{\dot{D_t}}^2= \frac{D_t^4}{C^2} \left[ \frac{2 \kappa \rho}{D_t(D_t-1)}
-k {\delta}^{-2} e^{-2C/D_t} \right].
\end{equation}
The continuity equation of the model universe with variable space
dimension
can be obtained by (\ref{16}) and (\ref{17})
\begin{equation}
\label{19}
\frac{d}{dt} \left[ \rho \left( \frac{a}{a_0} \right)^{D_t} V_{D_t} \right]
+ p \frac{d}{dt} \left[ \left( \frac{a}{a_0} \right )^{D_t} V_{D_t} \right] =0.
\end{equation}
All above equations, from (\ref{1}) to (\ref{19}), are written and
discussed in Refs.\cite{2}-\cite{9}. Let us here calculate the
present value of the deceleration parameter $q_0$ in this model.
We use (\ref{17}) in the case of flat universe, i.e. $k=0$.
Therefore, we get
\begin{eqnarray}
\label{20}
q_0 & \equiv & \frac{{-\ddot a}\,{a}}{{\dot a}^2} \bigg|_0\nonumber\\
    &  =     & -\frac{D_0^2}{2C}\frac{d \ln V_D}{dD} \bigg|_{D=D_0}-1
-\frac{D_0(2D_0-1)}{2C(D_0-1)}\nonumber\\
& + & \frac{D_0}{2}-\frac{p_0D_0^2}{2\rho_0} \left( \frac{1}{C}
\frac{d \ln V_D}{dD} \bigg|_{D=D_0} - \frac{1}{D_0} \right),
\end{eqnarray}
where $p_0$ and $\rho_0$ are the present value of the pressure and
the energy density respectively, satisfying the present equation of
state in the model
\begin{equation}
\label{21}
p_0=\omega_0 \rho_0.
\end{equation}
Here $\omega_0$ is the parameter of the equation of state at the present
time. Therefore we can rewrite (\ref{20}) in terms of $\omega_0$ as
\begin{equation}
\label{22}
q_0=\frac{D_0^2}{2} \left( 1+\omega_0 \right) \left( -\frac{1}{C}
\frac{d \ln V_D}{d D} \bigg|_{D=D_0}
+ \frac{1}{D_0} \right) - 1 - \frac{D_0(2D_0 -1)}{2C(D_0-1)}.
\end{equation}
This equation can be directly obtained by the time derivative of
(\ref{16}) in the case of a flat universe.

Concerning the dynamics in a flat universe, the authors of Ref.\cite{10}
conclude on the validity of the Cosmic
Concordance version of the $\Lambda$CDM Model that is
$\Omega_M \approx 0.3$, $\Omega_{\Lambda} \approx 0.7$,
$\omega(z=0) \approx -1$ and no rapid evolution of the equation of state
parameter. 
In the limit of constant space dimension, i.e. $C \to +\infty$,
from Eq.(\ref{22}) the deceleration parameter today is given by
\begin{equation}
\label{23}
q_0=\frac{D_0}{2} \left( 1+\omega_0 \right) - 1.
\end{equation}
Taking $\omega_0 \approx -1$ and $D_0=3$, the value of $q_0$ in the
universe with constant space dimension satisfies $q_0 \approx -1$.
In the case of decrumpling model, by considering $\omega_0 \approx -1$
and $D_0=3$,
the value of the deceleration parameter satisfies
\begin{equation}
\label{24}
q_0 \approx - 1 - \frac{15}{4C},
\end{equation}
(see Eq.(\ref{22}) and the values of $C$ given in Table 1).
Therefore, decrumpling model explains the universe is accelerating
today, i.e. $q_0 < 0$  or ${\ddot a} > 0$ and in this model
the reason for the present cosmic acceleration of the universe is due
to the cosmological constant which is the simplest candidate of the dark
energy. In the case of matter-dominated universe, i.e. $p=0$, we do not
analyze Eq.(\ref{22}) because supernovae searches have shown that a
simplest matter-dominated and decelerating universe should be ruled out.

\section*{Acknowledgments}
F.N. thanks Amir and Shahrokh for useful helps and also
thanks her birthplace city, Neishabour in Iran, where she
researches in physics and cosmology.


\begin{thebibliography}{99}
\bibitem{1} For a review of the observational evidence for the
acceleration of the expansion of the universe, see, e.g. N.A. Bahcall,
J.P. Ostriker, S. Perlmutter, and P.J. Steinhardt, Science 284 (1999)
1481, astro-ph/9906463.
\bibitem{2} F. Nasseri, Int.J.Mod.Phys.D 14 (2005) 1577, hep-th/0412070.
\bibitem{3} F. Nasseri and S.A. Alavi, Int. J. Mod. Phys. D 14 (2005) 621,
hep-th/0410259.
\bibitem{4} F. Nasseri and S. Rahvar, Mod. Phys. Lett. A 20 (2005) 2467,
astro-ph/0212371.
\bibitem{5} F. Nasseri, Phys. Lett. B 538 (2002) 223, gr-qc/0203032.
\bibitem{6} F. Nasseri and S. Rahvar, Int. J. Mod. Phys. D 11 (2002) 511,
gr-qc/0008044.
\bibitem{7} F. Nasseri, Ph.D thesis, Sharif University of Technology,
Tehran, Iran, 1999.
\bibitem{8} R. Mansouri, F. Nasseri and M. Khorrami, Phys.Lett.A 259
(1999) 194, gr-qc/9905052.
\bibitem{9} R. Mansouri and F. Nasseri, Phys. Rev. D 60 (1999) 123512,
gr-qc/9902043.
\bibitem{10} A.G. Riess et al., Astrophys.J. 607 (2004) 665,
astro-ph/0402512.
\end{thebibliography}
\end{document}